\documentclass[conference,10pt]{IEEEtran}
\IEEEoverridecommandlockouts

\normalsize

\usepackage[T1]{fontenc}
\usepackage{amsfonts}
\usepackage{amsmath}
\usepackage{amsthm}
\usepackage{amssymb}
\usepackage{caption}
\usepackage[labelformat=simple]{subcaption}
\usepackage{cite,graphicx,algorithm}
\usepackage{algorithmic}
\usepackage{color}
\usepackage{bm}

\ifCLASSINFOpdf
\else
\fi

\theoremstyle{plain}

\begin{document}
%
\title{Structured-Sparsity-Aware Joint User Activity Detection and Channel Estimation for OTFS-Based Grant-Free Random Access\vspace{-0.36em}}
%
%
%

\author{
\IEEEauthorblockN{Yao~Ge$^{1}$, Yirui~Luo$^{1}$, Yuhao~Chi$^{2}$, Yufei~Zhao$^{1}$, Yong~Liang~Guan$^{1}$, David~Gonz\'{a}lez~G.$^{3}$, and Zhi~Ding$^{4}$}
\IEEEauthorblockA{$^{1}$AUMOVIO-NTU Corporate Lab, Nanyang Technological University, Singapore}
\IEEEauthorblockA{$^{2}$State Key Laboratory of Integrated Services Networks, Xidian University, China}
\IEEEauthorblockA{$^{3}$Wireless Communications Technologies Group, AUMOVIO Germany GmbH}
\IEEEauthorblockA{$^{4}$Department of Electrical and Computer Engineering, University of California at Davis, Davis, CA 95616 USA}
\IEEEauthorblockA{Emails: $^{1}$\{yao.ge;yufei.zhao;eylguan\}@ntu.edu.sg; yirui001@e.ntu.edu.sg, $^{2}$yhchi@xidian.edu.cn, \\$^{3}$david.gonzalez.g@ieee.org, $^{4}$zding@ucdavis.edu}
}%

%
%

\markboth{}%
{}
%



\maketitle

\begin{abstract}
Grant-free random access (GFRA) is a promising solution for massive machine-type communications (mMTC) in future wireless networks. However, reliable user activity detection and channel estimation are critical challenges, particularly when orthogonal time–frequency space (OTFS) modulation is integrated with GFRA to address doubly selective channels induced by high mobility. In this paper, we propose an OTFS-based GFRA framework that exploits the inherent structured sparsity of delay–Doppler channels. By adopting a basis expansion model (BEM), we formulate joint user activity detection and channel estimation as a structured compressive sensing problem. A bi-level sparsity structure is identified, consisting of common sparsity across multiple receive antennas and activation sparsity across mMTC users. To effectively leverage this structure, we construct a two-layer factor graph and develop a structured sparsity expectation propagation (SS-EP) algorithm for efficient Bayesian inference. Simulation results demonstrate that the proposed scheme significantly outperforms existing benchmarks.
\end{abstract}

\begin{IEEEkeywords}
Grant-free random access, orthogonal time frequency space (OTFS), massive machine-type communications, compressed sensing, structured sparsity.
\end{IEEEkeywords}

%
\IEEEpeerreviewmaketitle

\section{Introduction}
%
%
%
%

As communication scenarios evolve toward massive machine-type communications (mMTC) dominated by large-scale Internet of Things (IoT) deployments, terminal access exhibits new characteristics, including massive connectivity, bursty traffic, and sporadic activity patterns \cite{9537931}. In this context, the grant-based four-step handshake procedure \cite{9097306} incurs prohibitive signaling overhead and access latency, rendering it unsuitable for supporting ultra-large-scale connectivity requirements. Therefore, grant-free random access (GFRA) has emerged as a promising access mechanism tailored for enabling low-latency and high-efficiency massive connectivity scenarios \cite{9060999}. 
Existing GFRA studies mainly focus on non-orthogonal schemes, where non-orthogonal preambles or signal superposition enable a larger number of simultaneous users under limited resources. However, these schemes require the receiver to jointly perform active user detection and channel estimation without prior knowledge of the active user set, thereby posing significant challenges for receiver design.

To address the challenges of jointly estimation processing in non-orthogonal GFRA, extensive efforts \cite{9097306,10329933} have been devoted to sparse signal recovery–based methods, iterative joint receiver designs, and learning-based approaches. With the evolution of wireless communication systems toward broadband transmission, the integration of GFRA with orthogonal frequency-division multiplexing (OFDM) is particularly attractive for supporting massive connectivity in multipath frequency-selective fading channels \cite{9725260,9586569}. 
These works establish a foundation for OFDM-based GFRA frameworks, but largely rely on quasi-static channel assumptions that may not hold in high-mobility doubly selective fadings, where strong Doppler effects destroy the subcarrier orthogonality in OFDM systems, resulting in severe performance loss.

To overcome the inherent limitations of OFDM in high-mobility scenarios, the integration of GFRA with Doppler-resilient waveforms \cite{11428186} has emerged as a key approach for enabling reliable large-scale multiple access.  
Orthogonal time frequency space (OTFS) modulation \cite{10891132}, due to its delay–Doppler domain signal representation, is robust to Doppler effects and can transform rapidly time-varying frequency-selective fading channels into quasi-static ones, making it particularly well suited for high-mobility GFRA scenarios. Existing works have explored the integration of GFRA with OTFS modulation for massive IoT access in low Earth orbit (LEO) satellite systems \cite{9849120,10959036}. \cite{11184438} and \cite{10680158} further explored OTFS-based GFRA in cell-free architectures and cooperative multi-satellite scenarios, leveraging sparsity-aware modeling and Bayesian learning techniques to support reliable user detection and channel estimation under high mobility. Despite the growing interest in OTFS-based GFRA, most existing works do not effectively exploit the structured sparsity of the delay–Doppler channel representation, resulting in performance loss and increased pilot overhead.

In this work, we propose an efficient OTFS-based GFRA framework for high-mobility scenarios. By leveraging a low-dimensional basis expansion model (BEM), we formulate joint user activity detection and channel estimation as a structured compressive sensing problem. We reveal and systematically exploit a previously unaddressed bi-level sparsity structure by developing a two-layer factor graph and proposing an efficient structured sparsity expectation propagation (SS-EP) algorithm to enhance the performance of joint estimation. Simulation results verify that the proposed SS-EP algorithm outperforms the existing benchmark schemes.

\section{System Model}\label{II_model}
We consider an uplink communication process in a GFRA system, where $K$ single-antenna vehicle users transmit to a base station (BS) equipped with $U$ antennas simultaneously. Due to the sporadic traffic pattern, only
${K_a}$ (${K_a} \ll K$) vehicle users are active during each channel coherence block. Let ${\alpha _k} \in \left\{ {0,1} \right\}$ denote the binary activity indicator of user $k$, i.e., ${\alpha _k} = 1$ indicates that user $k$ is active and vice versa. A classical two-phase GFRA scheme is developed, consisting of a pilot transmission phase with $L$ symbols, followed by a data transmission phase. Each user is preassigned a unique but not necessarily orthogonal pilot sequence ${{\bf{x}}_k} = {\left[ {{x_{k,1}},{x_{k,2}}, \cdots ,{x_{k,L}}} \right]^T} \in {\mathbb{C}^{L \times 1}}$ with ${x_{k,\ell}} \sim \mathcal{CN}\left( {0,{1 \mathord{\left/
 {\vphantom {1 {\sqrt L }}} \right.
 \kern-\nulldelimiterspace} {\sqrt L }}} \right),k = 1,2, \cdots ,K,\;\ell = 1,2, \cdots ,L$. This work focuses on the pilot transmission phase, where the active users send specific pilot sequences to BS for joint user activity detection and channel estimation. Each user employs an OTFS-based transmission scheme to effectively combat the doubly-selective fading effects.

Specifically, the pilot sequences ${{\bf{x}}_k} \in {\mathbb{C}^{L \times 1}}$ are arranged into the two-dimensional delay-Doppler (DD) domain grids ${{\bf{X}}_k} \in {\mathbb{C}^{M \times N}}$, i.e., ${{\bf{X}}_k} = \text{invec}({{\bf{x}}_k})$, where $\text{invec}(  \cdot  )$ denotes the inverse vectorization of a vector. Here, $M$ and $N$ are the number of discretization bins in the OTFS delay and Doppler domains, respectively.\footnote{For simplicity, we assume that the pilot length $L=MN$.} Then, the transmitted time domain pilot signal ${{\bf{s}}_k} \in {\mathbb{C}^{L \times 1}}$ is obtained by first applying the inverse symplectic finite Fourier transform (SFFT) to the DD domain signal matrix ${{\bf{X}}_k}$, followed by the Heisenberg transform using a rectangular pulse, i.e., 
\begin{align}
{{\bf{s}}_k} = \text{vec}\left\{ {{\bf{F}}_M^H\left( {{{\bf{F}}_M}{{\bf{X}}_k}{\bf{F}}_N^H} \right)} \right\} = \left( {{\bf{F}}_N^H \otimes {{\bf{I}}_M}} \right){{\bf{x}}_k},
\end{align}
where $ \otimes $ and $\text{vec}(  \cdot  )$ denote the Kronecker product operator and the column-wise vectorization of a matrix, respectively. ${{\bf{F}}_M} \in {\mathbb{C}^{M \times M}}$ and ${{\bf{F}}_N} \in {\mathbb{C}^{N \times N}}$ are the normalized fast Fourier transform (FFT) matrices of size $M$ and $N$, respectively.

To overcome the inter-frame interference, we then append a cyclic prefix (CP) in front of the generated time domain pilot signal for each user, where the length of CP is no shorter than the largest channel delay spread across all users. The resulted time domain pilot signal of each active user is finally sent out to the BS through the doubly-selective fading channels.

The time-delay domain equivalent channel matrix between $k$-th user and $u$-th BS antenna ${{\bf{H}}_{uk}} \in {\mathbb{C}^{L \times L}}$ is defined in a circular shift form, analogous to ${{\bf{H}}_{ch}}$ in (45) of \cite{10891132}, 
where its entries ${h_{uk}}\left[ {c,p} \right],c = 0,1, \cdots ,L - 1,p = 0,1, \cdots {P_{k}} - 1$ denotes the channel response of the $p$-th tap at $c$-th time instant. The value of 
${{P_{k}}}$, representing the maximum channel tap between the $k$-th user and the BS.
Considering Jakes time-varying channel model, the time-correlated function of $p$-th tap is defined as 
\begin{align}
\mathbb{E}\left\{ {{h_{uk}}\left[ {c,p} \right]h_{uk}^*\left[ {c',p} \right]} \right\} = \sigma _p^2{J_0}\left( {2\pi {f_{\max }}{T_s}\left( {c - c'} \right)} \right),
\end{align}
where $\sigma _p^2$ is the power gain of $p$-th tap and ${J_0}\left(  \cdot  \right)$ refers to the zeroth-order Bessel function of the first kind. The parameters ${{f_{\max }}}$ and ${{T_s}}$ denote the maximum Doppler frequency and the sampling interval, respectively. For simplicity, BEM has been extensively utilized to represent time-varying wireless channels by expressing the channel coefficients as a linear combination of a set of predefined basis functions, weighted by corresponding expansion coefficients. Thus, the channel impulse response of $p$-th tap at $c$-th time instant between $k$-th user and $u$-th BS antenna can be characterized as
\begin{align}
{h_{uk}}\left[ {c,p} \right] = \sum\limits_{q = 0}^Q {c_{q,p}^{u,k}{e^{j{\omega _q}c}}}.
\end{align}
Here, $Q = 2\left\lceil {N{{\bar f}_{\max }}} \right\rceil$ denotes the order of basis functions used in the BEM, where ${{\bar f}_{\max }} = {{{f_{\max }}} \mathord{\left/
 {\vphantom {{{f_{\max }}} {\Delta f}}} \right.
 \kern-\nulldelimiterspace} {\Delta f}}$ is the normalized maximum Doppler
frequency with ${\Delta f}$ being subcarrier spacing. The modeling frequency of the $q$-th BEM basis function is given by ${\omega _q} = \frac{{2\pi }}{L}\left( {q - \left\lceil {\frac{Q}{2}} \right\rceil } \right)$, and ${c_{q,p}^{u,k}}$ denotes the $q$-th BEM coefficient of the $p$-th channel tap between the $k$-th user and $u$-th BS antenna. The operator $\left\lceil  \cdot  \right\rceil$ represents the ceiling (round-up) function.

Thus, the time-delay domain equivalent channel matrix ${{\bf{H}}_{uk}}$ can be further rewritten as
\begin{align}
{{\bf{H}}_{uk}} \!=\! \sum\limits_{q = 0}^Q {{{\bf{D}}_q}{\bf{C}}_q^{u,k}} 
\!=\! \sum\limits_{q = 0}^Q {{{\bf{D}}_q}{\bf{F}}_L^H\text{diag}\left\{ {{{\bf{F}}_{L \times {P_{k}}}}{\bf{c}}_q^{u,k}} \right\}{{\bf{F}}_L}},\label{H_BEM}
\end{align}
where ${{\bf{D}}_q} = \text{diag}\left\{ {1,{e^{j{\omega _q}}},{e^{j2{\omega _q}}}, \cdots ,{e^{j{\omega _q}(L - 1)}}} \right\}$ and ${\bf{C}}_q^{u,k} \in {\mathbb{C}^{L \times L}}$ is a circulant matrix,
which can be decomposed into
\begin{align}
{\bf{C}}_q^{u,k} = {\bf{F}}_L^H\text{diag}\left\{ {{{\bf{F}}_{L \times {P_{k}}}}{\bf{c}}_q^{u,k}} \right\}{{\bf{F}}_L},
\end{align}
where ${\bf{c}}_q^{u,k} = {\left[ {c_{q,0}^{u,k},c_{q,1}^{u,k}, \cdots ,c_{q,{P_{k}} - 1}^{u,k}} \right]^T}$ denotes the $q$-th BEM coefficient vector between the $k$-th user and $u$-th BS antenna. ${{\bf{F}}_L} \in {\mathbb{C}^{L \times L}}$ is the normalized $L$-point FFT matrix and ${{{\bf{F}}_{L \times {P_{k}}}}} \in {\mathbb{C}^{L \times {{P_{k}}}}}$ represents the submatrix consisting of the first ${{P_{k}}}$ columns of ${{\bf{F}}_L}$. 

After discarding the CP, the time domain received signal ${{\bf{r}}_u}\in {\mathbb{C}^{L \times 1}}$ at the $u$-th BS antenna can be expressed as
\begin{align}
{{\bf{r}}_u} = \sum\limits_{k = 1}^K {{\alpha _k}{{\bf{H}}_{uk}}{{\bf{s}}_k}}  + {{{\bm{\bar \omega }}}_u},
\end{align}
where ${{{\bm{\bar \omega }}}_u} \in {\mathbb{C}^{L \times 1}} \sim \mathcal{CN}\left( {{\bf{0}},{\sigma ^2}{{\bf{I}}_L}} \right)$ represents the complex additive white Gaussian noise (AWGN) at the $u$-th BS antenna with the noise power ${{\sigma ^2}}$.

The resulting time domain signal ${{\bf{r}}_u}\in {\mathbb{C}^{L \times 1}}$ is then devectorized into a matrix ${{\bf{R}}_u} \in {\mathbb{C}^{M \times N}}$, followed by the OTFS demodulation, i.e., Wigner transform and then the SFFT, 

\noindent yielding the DD domain signal ${{\bf{y}}_u} \in {\mathbb{C}^{L \times 1}}$ given by
\begin{align}
{{\bf{y}}_u} =& \text{vec}\left\{ {{\bf{F}}_M^H\left( {{{\bf{F}}_M}{{\bf{R}}_u}} \right){{\bf{F}}_N}} \right\} = \left( {{{\bf{F}}_N} \otimes {{\bf{I}}_M}} \right){{\bf{r}}_u}\nonumber\\
 = &\left( {{{\bf{F}}_N} \otimes {{\bf{I}}_M}} \right)\sum\limits_{k = 1}^K {{\alpha _k}{{\bf{H}}_{uk}}\left( {{\bf{F}}_N^H \otimes {{\bf{I}}_M}} \right){{\bf{x}}_k}}  + {{\bm{\omega }}_u},\label{y_DD}
\end{align}
where ${{\bm{\omega }}_u} = \left( {{{\bf{F}}_N} \otimes {{\bf{I}}_M}} \right){{{\bm{\bar \omega }}}_u}$.

By substituting (\ref{H_BEM}) into (\ref{y_DD}), we can obtain
\begin{align}
&{{\bf{y}}_u} = \sum\limits_{k = 1}^K {\sum\limits_{q = 0}^Q {{\alpha _k}\left( {{{\bf{F}}_N} \otimes {{\bf{I}}_M}} \right){{\bf{D}}_q}{\bf{F}}_L^H\text{diag}\left\{ {{{\bf{F}}_{L \times {P_k}}}{\bf{c}}_q^{u,k}} \right\}} } \nonumber\\
& \quad\quad\quad\times {{\bf{F}}_L}\left( {{\bf{F}}_N^H \otimes {{\bf{I}}_M}} \right){{\bf{x}}_k} + {{\bm{\omega }}_u}\nonumber\\
& = \sum\limits_{k = 1}^K {\sum\limits_{q = 0}^Q {{\alpha _k}{\bf{\Phi }}_q^k{\bf{c}}_q^{u,k}} }  + {{\bm{\omega }}_u}\nonumber\\
& = \sum\limits_{k = 1}^K {{\alpha _k}{{\bf{\Phi }}^k}{{\bf{c}}^{u,k}}}  + {{\bm{\omega }}_u},
\end{align}
where ${{\bf{\Phi }}^k} = \left[ {{\bf{\Phi }}_0^k,{\bf{\Phi }}_1^k, \cdots ,{\bf{\Phi }}_Q^k} \right] \in {\mathbb{C}^{L \times \left( {Q + 1} \right){P_k}}}$ and ${{\bf{c}}^{u,k}} = {\left[ {{{\left( {{\bf{c}}_0^{u,k}} \right)}^T},{{\left( {{\bf{c}}_1^{u,k}} \right)}^T}, \cdots ,{{\left( {{\bf{c}}_Q^{u,k}} \right)}^T}} \right]^T} \in {\mathbb{C}^{\left( {Q + 1} \right){P_k} \times 1}}$. Here, we define ${\bf{\Phi }}_q^k = \left( {{{\bf{F}}_N} \otimes {{\bf{I}}_M}} \right){{\bf{D}}_q}{\bf{F}}_L^H\text{diag}\left\{ {{{\bf{F}}_L}\left( {{\bf{F}}_N^H \otimes {{\bf{I}}_M}} \right){{\bf{x}}_k}} \right\}{{\bf{F}}_{L \times {P_k}}}$.

Finally, the received pilot signal in DD domain over $U$ BS antennas can be written as
\begin{align}
{\bf{Y}} = \sum\limits_{k = 1}^K {{\alpha _k}{{\bf{\Phi }}^k}{{\bf{C}}^k}}  + {\bf{W}}
 = {\bf{\Phi H}} + {\bf{W}},\label{CS_Model}
\end{align}
where ${\bf{Y}} = \left[ {{{\bf{y}}_1},{{\bf{y}}_2}, \cdots ,{{\bf{y}}_U}} \right] \in {\mathbb{C}^{L \times U}}$, ${{\bf{C}}^k} = \left[ {{{\bf{c}}^{1,k}},{{\bf{c}}^{2,k}}, \cdots ,{{\bf{c}}^{U,k}}} \right] \in {\mathbb{C}^{\left( {Q + 1} \right){P_k} \times U}}$ and ${\bf{W}} = \left[ {{{\bm{\omega }}_1},{{\bm{\omega }}_2}, \cdots ,{{\bm{\omega }}_U}} \right] \in {\mathbb{C}^{L \times U}}$. We also define ${{\bf{ \Phi }}} = \left[ {{{\bf{\Phi }}^1},{{\bf{\Phi }}^2}, \cdots ,{{\bf{\Phi }}^K}} \right] \in {\mathbb{C}^{{L \times P\left( {Q + 1} \right)}}}$ and ${\bf{H}} = {\left[ {{\alpha _1}{{\left( {{{\bf{C}}^1}} \right)}^T},{\alpha _2}{{\left( {{{\bf{C}}^2}} \right)}^T}, \cdots ,{\alpha _K}{{\left( {{{\bf{C}}^K}} \right)}^T}} \right]^T} \in {\mathbb{C}^{P\left( {Q + 1} \right) \times U}}$ with $P = \sum\limits_{k = 1}^K {{P_k}}$. Therefore, the joint user activity
detection and channel estimation problem is now formulated
as a generalized multiple measurement vector (MMV) compressed sensing problem with structured sparsity. To this end, we develop a two-layer factor graph and propose a novel SS-EP algorithm to enhance the performance of joint estimation, which will be detailed in the next section.

\section{Proposed SS-EP Algorithm}\label{III_MLEP}
\begin{figure}
  \centering
  \includegraphics[width=2.7in]{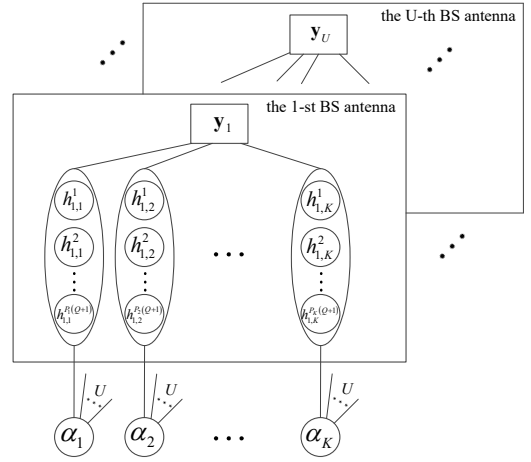}
  \caption{Factor graph of the factorized result in the equation (\ref{factor_result}). The circles and squares denote variable nodes and factor nodes, respectively.}\label{FG_model}
\end{figure}

To effectively leverage the prior information on structured sparsity, we introduce the proposed SS-EP algorithm in this section. The derivation starts with the posterior probability $p\left( {{\bf{H}}\left| {{\bf{Y}},{\bf{\Phi }}} \right.} \right)$, which can be expressed as
\begin{subequations}
\begin{align}
&p\left( {{\bf{H}}\left| {{\bf{Y}},{\bf{\Phi }}} \right.} \right)  \!= \! p\left( {{\bf{H}},{\bm{\alpha }}\left| {{\bf{Y}},{\bf{\Phi }}} \right.} \right)
 \!\propto \! p\left( {{\bf{Y}}\left| {{\bf{H}},{\bm{\alpha }},{\bf{\Phi }}} \right.} \right)p\left( {{\bf{H}},{\bm{\alpha }}} \right)\nonumber\\
& = \prod\limits_{u = 1}^U {p\left( {{{\bf{y}}_u}\left| {{{\bf{h}}_u},{\bm{\alpha }},{\bf{\Phi }}} \right.} \right)\prod\limits_{k = 1}^K {p\left( {{{\bf{H}}_k}\left| {{{\alpha }_k}} \right.} \right)p\left( {{{\alpha }_k}} \right)} } \\
& = \prod\limits_{u = 1}^U {p\left( {{{\bf{y}}_u}\left| {{{\bf{h}}_u},{\bm{\alpha }},{\bf{\Phi }}} \right.} \right)\prod\limits_{u = 1}^U {\prod\limits_{k = 1}^K {p\left( {{{\bf{h}}_{u,k}}\left| {{\alpha _k}} \right.} \right)p\left( {{\alpha _k}} \right)} } }, \label{factor_result}
\end{align}
\end{subequations}
where ${\bm{\alpha }} = {\left[ {{{\alpha }_1},{{\alpha }_2}, \cdots ,{{\alpha }_K}} \right]^T}$, ${{\bf{h}}_u} \in {\mathbb{C}^{P\left( {Q + 1} \right) \times 1}}$ is the $u$-th column of matrix ${\bf{H}}$, ${{\bf{H}}_k}={\alpha _k}{{\bf{C}}^k} \in {\mathbb{C}^{{P_k}\left( {Q + 1} \right) \times U}}$, and ${{\bf{h}}_{u,k}} \in {\mathbb{C}^{{P_k}\left( {Q + 1} \right) \times 1}}$ represents the $u$-th column of matrix ${{\bf{H}}_k}$. Fig. \ref{FG_model} illustrates the corresponding two-layer factor graph, based on which the SS-EP algorithm will be presented subsequently.

To address the MMV problem, the main idea is to decompose it into $U$ independent single measurement vector (SMV) subproblems, which can be solved in parallel for computational efficiency enhancement, i.e., (\ref{CS_Model}) can be decomposed into
\begin{align}
{{\bf{y}}_u} = {\bf{\Phi }}{{\bf{h}}_u} + {{\bm{\omega }}_u},\; u = 1,2, \cdots ,U.
\end{align}
In addition, our proposed SS-EP algorithm can exploit the prior information of structured sparsity and joint sparsity in the original MMV problem by exchanging soft information of the indicator vector ${{\alpha }_k}$ across different SMV subproblems. The two-layer factor graph of the SS-EP algorithm consists of three types of nodes, i.e., $U$ observation nodes ${{\bf{y}}_u},u = 1,2, \cdots ,U$; $UP\left( {Q + 1} \right)$ variable nodes $h_{u,k}^{i}$, $u = 1,2, \cdots ,U,i = 1,2, \cdots, {P_k}\left( {Q + 1} \right),k = 1,2, \cdots ,K$; and $K$ activity indicator nodes ${\alpha _{k}}$, $k = 1,2, \cdots ,K$. Here, the \textit{a priori} joint distribution of ${{\bf{h}}_{u,k}}$ is assumed to be
\begin{align}
&p\left( {{{\bf{h}}_{u,k}}\left| {{\alpha _k}} \right.} \right) = \prod\limits_{i = 1}^{{P_k}\left( {Q + 1} \right)} {p\left( {\left. {h_{u,k}^i} \right|{\alpha _k}} \right)} \nonumber\\
 & = \prod\limits_{i = 1}^{{P_k}\left( {Q + 1} \right)} {\left[ {\left( {1 - {\alpha _k}} \right)\delta \left( {h_{u,k}^i} \right) + {\alpha _k}\mathcal{CN}\left( {\gamma _{u,k}^i,\beta _{u,k}^i} \right)} \right]},\label{h_prior}
\end{align}
where ${h_{u,k}^{i}}$ is zero when the indicator ${{\alpha _{k}} = 0}$, and ${h_{u,k}^{i}}$ is belonging to a Gaussian distribution ${\mathcal{CN}\left( {\gamma _{u,k}^{i},\beta _{u,k}^{i}} \right)}$ with mean ${\gamma _{u,k}^{i}}$ and variance ${\beta _{u,k}^{i}}$ when the indicator ${{\alpha _{k}} = 1}$.

In addition, the \textit{a priori} distribution of activity indicator ${{\alpha _{k}}}$ can be expressed as 
\begin{align}\label{alpha_prior}
p\left( {{\alpha _{k}}} \right) = \left( {1 - {\lambda _{k}}} \right)\delta \left( {{\alpha _{k}}} \right) + {\lambda _{k}}\delta \left( {{\alpha _{k}} - 1} \right),
\end{align}
where ${{\lambda _{k}}}$ denotes the \textit{a priori} probability that ${{\alpha _{k}} = 1}$, i.e., the \textit{a priori} probability that the $k$-th user is active. 

\begin{algorithm}
\caption{Proposed SS-EP Algorithm}
\label{alg:A}
\begin{algorithmic}
\STATE {Input: ${\bf{Y}}$, ${\bf{\Phi }}$, ${\bm{\gamma }}$, ${\bm{\beta }}$, $\varepsilon$, $\varrho$ and ${I_{max}}$.}
\STATE {Initialization: ${{\bm{\mu }}_u^{\left( {0} \right)}}$, ${\bm{\eta }}_u^{\left( 0 \right)},u = 1,2, \cdots ,U$ and iteration count ${\ell  = 1}$.}
\REPEAT
\FOR[parallel computing]{$u = 1,2, \cdots ,U$}
\STATE {1)\; Observation node ${{\bf{y}}_u}$ generates the extrinsic mean $z_{u,m}^{\left( \ell  \right)}$ and variance $v_{u,m}^{\left( \ell  \right)}$ in (\ref{linear_ex}), then sends them to the variable nodes $h_{u,k}^{i}, i = 1,2, \cdots ,{P_k}\left( {Q + 1} \right),k = 1,2, \cdots ,K$;}
\STATE {2)\; Each variable node $h_{u,k}^{i}$ computes the probability message ${\bf{q}}_{u,k}^{i,\left( \ell  \right)}$ in (\ref{variab_q}), and transmits them to the connected activity indicator node ${{\alpha _{k}}}$;}
\ENDFOR
\STATE \quad{3)\; Each activity indicator node ${{\alpha _{k}}}$ computes the probability message ${\bm{\pi }}_{u,k}^{i,\left( \ell  \right)}$ in (\ref{post_pro}), and delivers them to the connected variable nodes $h_{u,k}^{i},u = 1,2, \cdots ,U,i = 1,2, \cdots ,{P_k}\left( {Q + 1} \right)$;}
\FOR[parallel computing]{$u = 1,2, \cdots ,U$}
\STATE {4)\; Each variable node $h_{u,k}^{i}$ generates the extrinsic mean $\mu _{u,m}^{\left( \ell  \right)}$ and variance $\eta _{u,m}^{\left( \ell  \right)}$ in (\ref{nonlinear_ex}), and passes them back to the connected observation node ${{\bf{y}}_u}$;}
\STATE {5)\; Compute the posteriori mean estimation ${\bf{\hat h}}_u^{\left( \ell  \right)},u = 1,2, \cdots ,U$ and the activity indicator variable estimation $\hat \alpha _k^{\left( \ell  \right)},k = 1,2, \cdots ,K$ in (\ref{est_h}) and (\ref{est_alpha});}
\ENDFOR
\STATE \quad{6)\; $\ell  = \ell  + 1$;}
\UNTIL{$\frac{{\left\| {{{{\bf{\hat H}}}^{\left( \ell  \right)}} - {{{\bf{\hat H}}}^{\left( {\ell  - 1} \right)}}} \right\|_F^2}}{{\left\| {{{{\bf{\hat H}}}^{\left( {\ell  - 1} \right)}}} \right\|_F^2}} < \varrho$ or $\ell={I_{max}}$.}
\STATE {Output: The estimated channel matrix ${{{{\bf{\hat H}}}^{\left( \ell  \right)}}}$, and the estimated user activity indicator $\hat \alpha _k^{\left( \ell  \right)},k = 1,2, \cdots ,K$.}
\end{algorithmic}
\end{algorithm}
The SS-EP algorithm is summarized in \textbf{Algorithm \ref{alg:A}}. We now detail the steps involved in iteration $\ell$:

1) From observation node ${{\bf{y}}_u}$ to variable nodes $h_{u,k}^{i}$, $i = 1,2, \cdots, {P_k}\left( {Q + 1} \right),k = 1,2, \cdots ,K$: For simplicity, the linear minimum mean squared error (LMMSE) criterion can be applied at the observation node to compute the \textit{a posteriori} estimation distribution of ${{\bf{h}}_u}$ with the covariance matrix and mean vector given as follows:
\begin{subequations}\label{LMMSE}
\begin{align}
{\bf{\bar V}}_u^{\left( \ell  \right)} &\!=\! {\left( {{\sigma ^{ - 2}}{{\bf{\Phi }}^H}{\bf{\Phi }} \!+\! \text{diag}{{\left\{ {{\bm{\eta }}_u^{\left( {\ell  - 1} \right)}} \right\}}^{ - 1}}} \right)^{ - 1}},\\
{\bf{\bar z}}_u^{\left( \ell  \right)} &\!=\! {\bf{\bar V}}_u^{\left( \ell  \right)}\left( {{\sigma ^{ - 2}}{{\bf{\Phi }}^H}{{\bf{y}}_u} \!+\! \text{diag}{{\left\{ {{\bm{\eta }}_u^{\left( {\ell  - 1} \right)}} \right\}}^{ - 1}}{\bm{\mu }}_u^{\left( {\ell  - 1} \right)}} \right),
\end{align}
\end{subequations}
where ${{\bm{\mu }}_u^{\left( {\ell  - 1} \right)}}$ and ${{\bm{\eta }}_u^{\left( {\ell  - 1} \right)}}$ are the \textit{a priori} mean and variance vectors of estimation ${{\bf{h}}_u}$, which can be obtained from the variable nodes in the ${\left( {\ell  - 1} \right)}$-th iteration. According to the Gaussian message
combining rule \cite{ge2021otfs,10050811}, we then update the extrinsic marginal distribution $q_E^{\left( \ell  \right)}\left( {{h_{u,m}}} \right) \sim \mathcal{CN}\left( {z_{u,m}^{\left( \ell  \right)},v_{u,m}^{\left( \ell  \right)}} \right)$, $m = 1,2, \cdots ,P\left( {Q + 1} \right)$ as follows:
\begin{subequations}\label{linear_ex}
\begin{align}
v_{u,m}^{\left( \ell  \right)} &= {\left[ {{{\left( {\bar V_{u,m}^{\left( \ell  \right)}} \right)}^{ - 1}} - {{\left( {\eta _{u,m}^{\left( {\ell  - 1} \right)}} \right)}^{ - 1}}} \right]^{ - 1}},\\
z_{u,m}^{\left( \ell  \right)} &= v_{u,m}^{\left( \ell  \right)}\left[ {\frac{{\bar z_{u,m}^{\left( \ell  \right)}}}{{\bar V_{u,m}^{\left( \ell  \right)}}} - \frac{{\mu _{u,m}^{\left( {\ell  - 1} \right)}}}{{\eta _{u,m}^{\left( {\ell  - 1} \right)}}}} \right],
\end{align}
\end{subequations}
where ${\bar V_{u,m}^{\left( \ell  \right)}}$ is the $m$-th diagonal entry of the covariance matrix ${\bf{\bar V}}_u^{\left( \ell  \right)}$. Then, the mean $z_{u,m}^{\left( \ell  \right)}$ and variance $v_{u,m}^{\left( \ell  \right)}$ are passed from observation node to the corresponding variable nodes. 

2) From variable nodes $h_{u,k}^{i},i = 1,2, \cdots ,{P_k}\left( {Q + 1} \right),u = 1,2, \cdots ,U$ to activity indicator node ${\alpha _{k}}$: At each variable node $h_{u,k}^{i}$, $z_{u,m}^{\left( \ell  \right)}$ is modeled as an AWGN observation of the real value $h_{u,k}^{i}$, i.e.,
\begin{align}\label{obser_model}
z_{u,m}^{\left( \ell  \right)} = h_{u,k}^{i} + \sqrt {v_{u,m}^{\left( \ell  \right)}} \tau ,
\end{align}
where $m = \bar P\left( {Q + 1} \right) + i$ and $\tau  \sim \mathcal{CN}\left( {0,1} \right)$ is independent of $h_{u,k}^{i}$. Here, we set $
\begin{aligned}
{\bar P}=
\begin{cases}
0,&k = 1\\
\sum\limits_{j = 1}^{k - 1} {{P_j}}, &k > 1
\end{cases}
\end{aligned}
$ for simplicity. 

Combining the observation model (\ref{obser_model}) with the \textit{a priori} distribution in (\ref{h_prior}), the estimate probability message $q_{u,k}^{i,\left( \ell \right)}\left( b \right)$ passed from variable node $h_{u,k}^{i}$ to activity indicator node ${\alpha _{k}}$ is given by
\begin{equation}\label{variab_q}
\begin{aligned}
q_{u,k}^{i,\left( \ell \right)}\left( b \right) &\buildrel \Delta \over = p\left( {{\alpha _{k}} = b\left| {h_{u,k}^{i}} \right.} \right) \\
 &\propto 
\begin{cases}
p\left( {z_{u,m}^{\left( \ell  \right)}\left| {h_{u,k}^{i} \ne 0} \right.} \right),&\text{if} \; b = 1,\\
p\left( {z_{u,m}^{\left( \ell  \right)}\left| {h_{u,k}^{i} = 0} \right.} \right), &\text{if} \; b = 0,
\end{cases}
\end{aligned}
\end{equation}
where $p\left( {z_{u,m}^{\left( \ell  \right)}\left| {h_{u,k}^{i} \ne 0} \right.} \right) = \frac{1}{{1 + \bar a_{u,m}^{\left( \ell  \right)}\exp \left( {\bar b_{u,m}^{\left( \ell  \right)}} \right)}}$ and $p\left( {z_{u,m}^{\left( \ell  \right)}\left| {h_{u,k}^{i} = 0} \right.} \right) = 1 - p\left( {z_{u,m}^{\left( \ell  \right)}\left| {h_{u,k}^{i} \ne 0} \right.} \right)$ with $\bar a_{u,m}^{\left( \ell  \right)} = \sqrt {\frac{{v_{u,m}^{\left( \ell  \right)} + \beta _{u,k}^{i}}}{{v_{u,m}^{\left( \ell  \right)}}}} $ and $\bar b_{u,m}^{\left( \ell  \right)} = \frac{1}{2}\left( {\frac{{{{\left( {z_{u,m}^{\left( \ell  \right)} - \gamma _{u,k}^{i}} \right)}^2}}}{{v_{u,m}^{\left( \ell  \right)} + \beta _{u,k}^{i}}} - \frac{{{{\left( {z_{u,m}^{\left( \ell  \right)}} \right)}^2}}}{{v_{u,m}^{\left( \ell  \right)}}}} \right)$, respectively. 

3) From activity indicator node ${{\alpha _{k}}}$ to variable nodes $h_{u,k}^{i},i = 1,2, \cdots ,{P_k}\left( {Q + 1} \right),u = 1,2, \cdots ,U$: According to the sum-product principle \cite{10159363}, the estimate \textit{a posteriori} probability message $\pi _{u,k}^{i,\left( \ell  \right)}\left( b \right)$ passed from activity indicator node ${{\alpha _{k}}}$ to variable node $h_{u,k}^{i}$ is given by
\begin{equation}\label{post_pro}
\begin{aligned}
&\pi _{u,k}^{i,\left( \ell  \right)}\left( b \right) = p\left( {{\alpha _{k}} = b\left| {{\bf{H}}_{k,\backslash \left\{ {\left( {u,i} \right)} \right\}}} \right.} \right)\\
 &\!\propto \!
\begin{cases}
{{\lambda _k}}{{\psi _k^{\left( \ell  \right)}\left( 1 \right)} \mathord{\left/
 {\vphantom {{\psi _k^{\left( \ell  \right)}\left( 1 \right)} {q_{u,k}^{d,i,\left( \ell  \right)}\left( 1 \right)}}} \right.
 \kern-\nulldelimiterspace} {q_{u,k}^{i,\left( \ell  \right)}\left( 1 \right)}},&\text{if} \; b\! =\! 1,\\
\left( {1 - {\lambda _k}} \right){{\psi _k^{\left( \ell  \right)}\left( 0 \right)} \mathord{\left/
 {\vphantom {{\psi _k^{\left( \ell  \right)}\left( 0 \right)} {q_{u,k}^{i,\left( \ell  \right)}\left( 0 \right)}}} \right.
 \kern-\nulldelimiterspace} {q_{u,k}^{i,\left( \ell  \right)}\left( 0 \right)}}, &\text{if} \; b\! =\! 0,
\end{cases}
\end{aligned}
\end{equation}
where ${{\bf{H}}_{k,\backslash \left\{ {\left( {u,i} \right)} \right\}}}$ represents ${{\bf{H}}_k}$ excluding ${h_{u,k}^{i}}$, and $\psi _k^{\left( \ell  \right)}\left( b \right)={\prod\limits_{u' = 1}^U {\prod\limits_{i' = 1}^{{P_k}\left( {Q + 1} \right)} {q_{u',k}^{i',\left( \ell  \right)}\left( b \right)} } }$. ${\lambda _{k}} \in \left[ {0,1} \right]$ denotes the \textit{a priori} probability that ${{\alpha _{k}} = 1}$, which is determined by empirical information or updated via the expectation maximization (EM) learning algorithm during iterations.

4) From variable nodes $h_{u,k}^{i}$, $i = 1,2, \cdots, {P_k}\left( {Q + 1} \right),k = 1,2, \cdots ,K$ to observation node ${{\bf{y}}_u}$: Combining the observation model (\ref{obser_model}) with the \textit{a priori} distribution in (\ref{h_prior}) as well as the \textit{a posteriori} probability in (\ref{post_pro}), we can obtain the approximate posterior distribution of ${h_{u,k}^{i}}$, which can be expressed as 
\begin{align}\label{post_h}
&p\left( {h_{u,k}^{i}\left| {z_{u,m}^{\left( \ell  \right)},{\alpha _{k}},{\bf{H}}_{k,\backslash \left\{ {\left( {u,i} \right)} \right\}}} \right.} \right)\nonumber\\ &\!\!\!=\!\! \left( {1 \!\!- \!\bar \lambda _{u,k}^{i,\left( \ell  \right)}} \right)\delta \left( {h_{u,k}^{i}} \right) \!+\! \bar \lambda _{u,k}^{i,\left( \ell  \right)}\mathcal{CN}\left( {\bar \gamma _{u,k}^{i,\left( \ell  \right)},\bar \beta _{u,k}^{i,\left( \ell  \right)}} \right),
\end{align}
where $\bar \lambda _{u,k}^{i,\left( \ell  \right)} = \frac{{\pi _{u,k}^{i,\left( \ell  \right)}\left( 1 \right)}}{{\pi _{u,k}^{i,\left( \ell  \right)}\left( 1 \right) + \pi _{u,k}^{i,\left( \ell  \right)}\left( 0 \right)}} \in \left[ {0,1} \right]$ denotes the approximate \textit{a posteriori} probability that ${h_{u,k}^{i} \ne 0}$. Following the Gaussian message combining rule \cite{ge2021otfs,10050811}, ${\bar \beta _{u,k}^{i,\left( \ell  \right)}}$ and ${\bar \gamma _{u,k}^{i,\left( \ell  \right)}}$ are respectively calculated by
\begin{align}
\bar \beta _{u,k}^{i,\left( \ell  \right)} = {\left[ {{{\left( {\beta _{u,k}^{i}} \right)}^{ - 1}} + {{\left( {v_{u,m}^{\left( \ell  \right)}} \right)}^{ - 1}}} \right]^{ - 1}},
\end{align}
and
\begin{align}
\bar \gamma _{u,k}^{i,\left( \ell  \right)} = \bar \beta _{u,k}^{i,\left( \ell  \right)}\left[ {\frac{{\gamma _{u,k}^{i}}}{{\beta _{u,k}^{i}}} + \frac{{z_{u,m}^{\left( \ell  \right)}}}{{v_{u,m}^{\left( \ell  \right)}}}} \right].
\end{align}

In accordance with (\ref{post_h}), the approximate posterior mean and variance of ${{\bf{h}}_u}$ can be formally expressed as
\begin{subequations}\label{nonlinear_post}
\begin{align}
&\bar \mu _{u,m}^{\left( \ell  \right)} = \bar \lambda _{u,k}^{i,\left( \ell  \right)}\bar \gamma _{u,k}^{i,\left( \ell  \right)},\\
\!\!\!\!\bar \eta _{u,m}^{\left( \ell  \right)} &\!=\!\! \left(\! {\bar \lambda _{u,k}^{i,\left( \ell  \right)} \!-\! {{\left( {\bar \lambda _{u,k}^{i,\left( \ell  \right)}} \!\right)}^2}} \!\right)\!\!{\left(\! {\bar \gamma _{u,k}^{i,\left( \ell  \right)}} \right)^2} \!\!+\! \bar \lambda _{u,k}^{i,\left( \ell  \right)}\bar \beta _{u,k}^{i,\left( \ell  \right)},
\end{align}
\end{subequations}
where $m = \bar P\left( {Q + 1} \right) + i$. We then update the extrinsic distribution $\bar q_E^{\left( \ell  \right)}\left( {{h_{u,m}}} \right) \sim CN\left( {\mu _{u,m}^{\left( \ell  \right)},\eta _{u,m}^{\left( \ell  \right)}} \right),m = 1,2, \cdots ,P\left( {Q + 1} \right)$, where 
\begin{subequations}\label{nonlinear_ex}
\begin{align}
\eta _{u,m}^{\left( \ell  \right)} &= {\left[ {{{\left( {\bar \eta _{u,m}^{\left( \ell  \right)}} \right)}^{ - 1}} - {{\left( {v_{u,m}^{\left( \ell  \right)}} \right)}^{ - 1}}} \right]^{ - 1}},\\
\mu _{u,m}^{\left( \ell  \right)} &= \eta _{u,m}^{\left( \ell  \right)}\left[ {\frac{{\bar \mu _{u,m}^{\left( \ell  \right)}}}{{\bar \eta _{u,m}^{\left( \ell  \right)}}} - \frac{{z_{u,m}^{\left( \ell  \right)}}}{{v_{u,m}^{\left( \ell  \right)}}}} \right].
\end{align}
\end{subequations}
Finally, ${{\bm{\mu }}_u^{\left( {\ell} \right)}}$ and ${{\bm{\eta }}_u^{\left( {\ell  } \right)}}$ are passed back to the observation node ${{\bf{y}}_u}$ to form the iterative loop.

5) Stopping criteria: At each variable node, the \textit{a posteriori} distribution of  ${h_{u,k}^{i}}$ can be expressed as
\begin{align}
&p\left( {h_{u,k}^{i}\left| {z_{u,m}^{\left( \ell  \right)},{\alpha _{k}}} \right.} \right) \nonumber\\ &\!\!=\! \left( {1 \!\!-\! \bar \pi _{u,k}^{i,\left( \ell  \right)}} \right)\!\delta\! \left( {h_{u,k}^{i}} \right) \!+\! \bar \pi _{u,k}^{i,\left( \ell  \right)}\mathcal{CN}\!\left( {\bar \gamma _{u,k}^{i,\left( \ell  \right)},\bar \beta _{u,k}^{i,\left( \ell  \right)}} \right),
\end{align}
where $\bar \pi _{u,k}^{i,\left( \ell  \right)} = \bar \pi _{k}^{\left( \ell  \right)} = \frac{{{\lambda _k}\psi _k^{\left( \ell  \right)}\left( 1 \right)}}{{{\lambda _k}\psi _k^{\left( \ell  \right)}\left( 1 \right) + \left( {1 - {\lambda _k}} \right)\psi _k^{\left( \ell  \right)}\left( 0 \right)}} \in \left[ {0,1} \right]$ denotes the \textit{a posteriori} probability that ${h_{u,k}^{i} \ne 0}$. Thus, the estimate posteriori mean of ${{h_{u,m}}}$ can be calculated by 
\begin{align}\label{est_h}
\hat h_{u,m}^{\left( \ell  \right)} = \bar \pi _{u,k}^{i,\left( \ell  \right)}\bar \gamma _{u,k}^{i,\left( \ell  \right)},
\end{align}
where $m = \bar P\left( {Q + 1} \right) + i,u = 1,2, \cdots ,U$.
The estimated activity indicator variable $\hat \alpha _k^{\left( \ell  \right)}$ of $k$-th user can be obtained as 
\begin{equation}\label{est_alpha}
\begin{aligned}
\hat \alpha _k^{\left( \ell  \right)}=
\begin{cases}
1,&\text{if} \; \bar \pi _k^{\left( \ell  \right)} > \varepsilon ,\\
0, &\text{otherwise},
\end{cases}
\end{aligned}
\end{equation}
where $\varepsilon $ is the user activity detection threshold that balances miss detection and false alarm rates.

The SS-EP algorithm terminates when either $\frac{{\left\| {{{{\bf{\hat H}}}^{\left( \ell  \right)}} - {{{\bf{\hat H}}}^{\left( {\ell  - 1} \right)}}} \right\|_F^2}}{{\left\| {{{{\bf{\hat H}}}^{\left( {\ell  - 1} \right)}}} \right\|_F^2}} < \varrho$ ($\varrho$ denotes the convergence threshold) or the maximum iteration number ${I_{max}}$ is reached. 

The proposed SS-EP algorithm effectively exploits the underlying structured sparsity priors to enhance the performance of joint user activity detection and channel estimation in OTFS-based GFRA systems. For each main loop iteration of the SS-EP algorithm, the main steps (\ref{LMMSE}), (\ref{linear_ex}), (\ref{variab_q}),  (\ref{post_pro}) and (\ref{nonlinear_ex}) have a complexity order $\mathcal{O}\left( {U\left[ {{{\left( {PD(Q + 1)} \right)}^3} + L{{\left( {PD(Q + 1)} \right)}^2} + PD(Q + 1)L + } \right.} \right.$ $\left. {\left. {2PD(Q + 1)} \right]} \right)$, $\mathcal{O}\left( {2PD(Q + 1)U} \right)$, $\mathcal{O}\left( {10PD(Q + 1)U} \right)$, $\mathcal{O}\left( {D{{\left( {P\left( {Q + 1} \right)U} \right)}^2}} \right)$ and $\mathcal{O}\left( {6PD(Q + 1)U} \right)$, respectively. Therefore, the overall complexity order is $\mathcal{O}\left( {U\left[ {{{\left( {PD(Q + 1)} \right)}^3} + L{{\left( {PD(Q + 1)} \right)}^2} + } \right.} \right.$ $\left. {\left. {UD{{\left( {P(Q + 1)} \right)}^2} \!+\! PD(Q + 1)L \!+\! 20PD(Q + 1)} \right]} \right)$ for the proposed SS-EP algorithm at each iteration.

\section{Simulation Results}\label{V_simulation}

In this section, we test the performance of our proposed joint estimation scheme for OTFS-based GFRA. Unless otherwise specified, we consider a carrier frequency centered at $4$ GHz with a subcarrier spacing of $15$ kHz. The delay–Doppler plane is configured with $M=32$ and $N=16$, leading to a pilot length $L=512$. The pilot sequences for each user are generated according to a random Gaussian distribution. The system accommodates $K = 200$ potential users, of which ${K_a} = 10$ users are active. Additionally, we adopt the typical 5G TDL-B channel model \cite{channel2017} with an exponential power delay profile. The BEM \cite{720248,10818747} is applied to characterize the time selective fading channels with maximum user velocity set to $150$ km/h. We assume that the BS equipped with $U=2$ antennas. 
The convergence threshold of the proposed algorithms and the maximum number of iterations are given by $\varrho  = {10^{ - 4}}$ and ${I_{max}} = 20$. 
The estimation performance is quantified by the normalized mean
square error (NMSE), defined as $\frac{{\left\| {{\bf{\hat H}} - {\bf{H}}} \right\|_F^2}}{{\left\| {\bf{H}} \right\|_F^2}}$. 

\begin{figure*}
\begin{subfigure}{0.33\textwidth}
  \centering
  \includegraphics[width=1.1\linewidth]{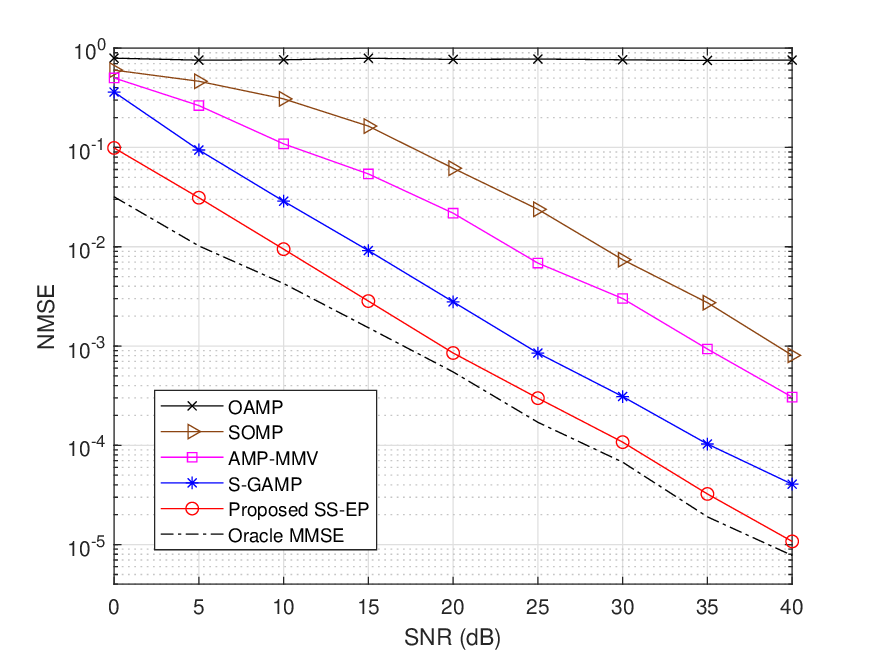}
  \caption{NMSE versus SNR.}
  \label{NMSE_compare}
\end{subfigure}
\begin{subfigure}{0.33\textwidth}
  \centering
  \includegraphics[width=1.1\linewidth]{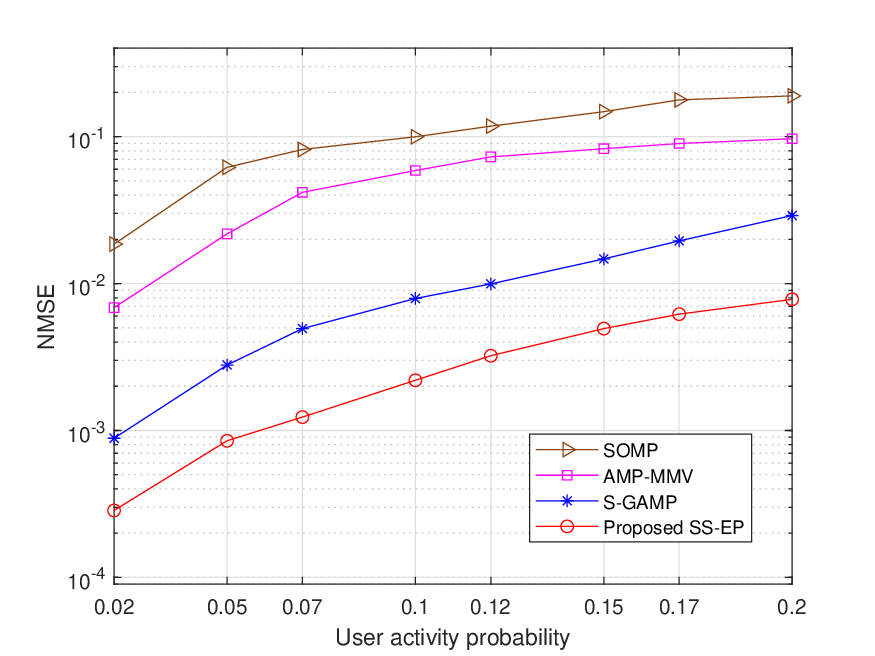}
  \caption{NMSE versus user activity probability.}
  \label{NMSE_UAP_compare}
\end{subfigure}
\begin{subfigure}{0.33\textwidth}
  \centering
  \includegraphics[width=1.1\linewidth]{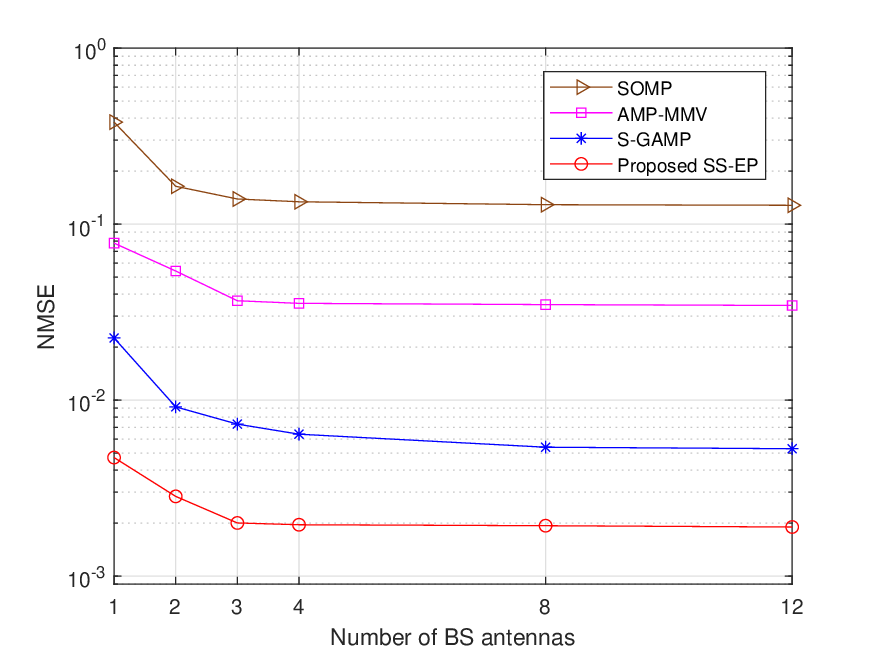}
  \caption{NMSE versus the number of BS antennas.}
  \label{NMSE_BS_compare}
\end{subfigure}
\caption{Performance test of the proposed joint estimation scheme for OTFS-based GFRA systems.}
\label{Perfor_compare}
\vspace{-1.5em}
\end{figure*}

In Fig. \ref{NMSE_compare}, we compare the NMSE performance of joint estimation with signal-to-noise ratio (SNR) for different estimation algorithms. It is observed that the orthogonal approximate message passing (OAMP) algorithm \cite{ge2021otfs} breaks down in the considered severely overloaded systems, owing to its inability to exploit the prior information on structured sparsity. The performance of the simultaneous orthogonal matching pursuit (SOMP) \cite{10448402}, approximate message passing multiple measurement vector (AMP-MMV) \cite{8264818} and structured generalized approximate message passing (S-GAMP) \cite{9586569} algorithms is progressively enhanced with increasing SNR, since they are able to partially exploit prior information, namely block-row sparsity. In contrast, our proposed SS-EP algorithm outperforms other benchmark schemes by fully accounting for the structured sparsity imposed by user activity, and approaches the performance of the Oracle minimum mean square error (MMSE) baseline, where the accurate active user index set is perfectly known at the BS. Without loss of generality, the OAMP estimation algorithm will no longer be considered in the subsequent simulations.

Fig. \ref{NMSE_UAP_compare} presents the joint estimation performance comparison for different estimation algorithms with different values of user activity probability (defined as ${K_a}/K$) under $\text{SNR}=20$ dB. As shown in Fig. \ref{NMSE_UAP_compare}, the performance of joint estimation for different algorithms decrease with the increase of the user activity probability. This is due to the fact that the system sparsity deteriorates as the user activity probability increases. However, our proposed SS-EP algorithm consistently achieves superior performance over the other benchmark schemes across various user activity probabilities. 

Fig. \ref{NMSE_BS_compare} shows the joint estimation performance comparison for different estimation algorithms versus the number of BS antennas under $\text{SNR}=15$ dB. It is observed that the proposed algorithm outperforms the baseline schemes, and their joint estimation performance initially improves and then saturates as the number of BS antennas increases. This benefit arises from the additional joint structured sparsity information provided by the increasing number of BS antennas.

\section{Conclusion}\label{VI_conclusion}
This paper proposed an OTFS-based GFRA framework for high-mobility mMTC systems. By leveraging a low-dimensional BEM, we formulated the joint user activity detection and channel estimation problem as a structured compressive sensing task in the delay–Doppler domain. A bi-level sparsity structure was identified and exploited through a two-layer factor graph and the proposed SS-EP algorithm. Simulation results demonstrated that the proposed scheme achieves superior performance compared to existing benchmarks.


%



\section*{Acknowledgment}
\vspace{-0.4em}
This work was supported by A*STAR under the RIE2025 Industry Alignment Fund–Industry Collaboration Projects (IAF-ICP) Funding Initiative (Award: I2501E0045), as well as cash and in-kind contribution from the industry partner(s). The work of Yuhao Chi was supported in part by the NSFC under Grant 62571399 and Grant 62471357, and in part by the Fundamental Research Funds for the Central Universities (QTZX26126).


\ifCLASSOPTIONcaptionsoff
  \newpage
\fi



%
%




\bibliographystyle{IEEEtran}
\footnotesize
\bibliography{ref_OTFS_CoMP}

%




\end{document}